\documentclass[aps,preprint]{revtex4}
\usepackage{amsmath}
\usepackage{epsfig}
\usepackage{float}
\usepackage{dcolumn} 

\begin{document}
\title{\bf Accidental crossings of eigenvalues in the one-dimensional complex PT-symmetric Scarf-II potential}  
\author{Zafar Ahmed$^1$, Dona Ghosh$^2$, Joseph Amal Nathan$^3$, Gaurang Parkar$^{4}$}
\affiliation{$~^1$Nuclear Physics Division, Bhabha Atomic Research Centre, Mumbai 400 085, India \\
$~^2$Astavinayak Society, Vashi, Navi-Mumbai, 400703, India \\
$~^3$Reactor Physics Design Division, Bhabha Atomic Research Centre, Mumbai 400085, India\\
$~^4$National Institute of Technology, Tiruchirappalli, Tamil Nadu 620015, India}
\email{1:zahmed@barc.gov.in, 2: rimidonaghosh@ gmail.com, 3:josephan@barc.gov.in, gaurangparkar@gmail.com}
\date{\today}
\begin{abstract}
\noindent
So far, the well known two branches of real discrete spectrum of complex PT-symmetric Scarf II potential are kept isolated. Here, we suggest that these two need to be brought together as doublets: $E^n_{\pm}(\lambda)$ with $n=0,1,2...$. Then if strength $(\lambda)$ of the imaginary part of the potential is varied smoothly some pairs of  real eigenvalue curves  can intersect and cross each other at $\lambda=\lambda_{*}$; this is unlike one dimensional Hermitian potentials. However, we show that the corresponding eigenstates
at $\lambda=\lambda_{*}$ are identical or linearly dependent denying degeneracy in one dimension, once again. Other pairs of eigenvalue curves  coalesce to complex-conjugate pairs completing the scenario of spontaneous breaking of PT-symmetry at $\lambda=\lambda_{c}$. To re-emphasize, sharply at $\lambda=\lambda_{*}$ and $\lambda_{c}$, two real eigenvalues coincide, nevertheless their corresponding eigenfunctions become identical or linearly dependent and the Hamiltonian looses diagonalizability.
\end{abstract}
\maketitle
In PT-symmetric quantum mechanics [1,2], one considers non-real, non-Hermitian  Hamiltonians which are invariant under the joint action of Parity ($P: x \rightarrow -x$) and Time-reversal ($T: i \rightarrow -i)$. Even the most simple Hamiltonian $H=p^2+V(x)$ corresponding to the Schr{\"o}dinger equation for these potentials has given amazing results. Based on numerical computations Bender and Boettcher [1] conjectured that the spectrum of $V_{BB}(x,\epsilon)=x^2(ix)^\epsilon$ was entirely real when $\epsilon \ge 0$. Next, for $-1< \epsilon < 0$ the spectrum consisted of a few real and the rest as complex-conjugate pairs of discrete eigenvalues. In the former case the energy eigenstates were also the eigenstates [1] of PT and the PT-symmetry was exact or unbroken.  Interestingly, $V_{BB}(x,2)=-x^4$ is a real Hermitian barrier (not a well), the real positive discrete spectrum has been aptly interpreted [3] as the reflectivity zeros in scattering from such potentials as $V(x)=-x^{2n+2}$, $n=1,2,3..$

Several  exactly solvable potentials were complexified to produce [4] exactly solvable
complex PT-symmetric potentials having real discrete
spectrum. Existence of two branches of real discrete
spectrum in complex PT-symmetric Scarf II  was revealed [5] and interpreted in terms of (unknown) quasi-parity [6]. Complex PT-symmetric Scarf II 
\begin{equation}
V(x)=-V_1 \mbox{sech}^2x + i |V_2| \mbox{sech}x \tanh x,~~V_1>0, V_2 \in {\cal R},
\end{equation}
was shown to be an exactly solvable model displaying the spontaneous breaking of PT-symmetry when the the strength of the imaginary part, $|V_2|$, exceeded a critical value of $V_{2c}=$ $V_1+1/4$ ($2\mu=1=\hbar^2$) [7], where $-V_1(V_1>0)$ was the strength of the real part. Such a phase transition of eigenvalues from real to complex conjugate  pairs in complex Scarf II has inspired 
several theoretical investigations [7-11]. On the other hand very interesting experiments in wave propagation and optics [12] have been performed where they realize PT-symmetry as an equal gain and loss medium. 

The paradigm model $V_{BB}(x,\epsilon)$ [1-2] of complex PT-symmetric potential was solved numerically for energy eigenvalues and the obtained eigenvalues can be seen  as $E_0,(E_1, E_2),(E_3, E_4),(E_5, E_6), ...$ in the increasing order. Only $E_0$ was unpaired, other levels were paired as doublets. These pairs coalesce  to complex conjugate pairs of eigenvalues in the parametric domain $\epsilon \in (-1,0)$. For $\epsilon \ge 0$, they open up to diverge from each other. In another model $V_A(x)=x^4+iAx$ [13] when $|A|<3.169$, the whole discrete spectrum is real and the eigenvalues can be arranged as $(E_0, E_1),(E_2, E_3),(E_4, E_5)...$, here the pairing of doublets unlike the case of $V_{BB}(x)$ starts from the ground state itself. In Hermitian case, doublets mean even(odd) parities or even(odd) numbers of nodes of eigenstates. In complex PT-symmetric quantum mechanics  the energy eigenfunctions are complex and they loose definite parity [14] hence the number of nodes are no more meaningful. The interesting question arising here is whether one can assign a quasi-parity to the levels/states of these doublets.

On the contrary, for the exactly solvable versatile complex PT-symmetric Scarf II, two separate branches of real discrete spectrum occur by virtue of the two linearly independent analytic solutions of the Schr{\"o}dinger equation. Here, one can have $E^n_+$ and $E^n_-$ with $\pm$ denoting some (unknown) quasi-parity [5,6]. Consequently, due to the isolation of these two branches several interesting features have gone un-noticed. The question arising here is as to what happens if these two branches are brought together.

In Hermitian quantum mechanics in two or more dimensions if two eigenvalues are equal but their corresponding eigenfunctions are different(not linearly dependent), this phenomenon is called degeneracy. In other words if a parameter in the Hamiltonian is varied slowly and smoothly two energy eigenvalue curves may come very close and avoid crossing each other or they can intersect to cross each other.  A degeneracy (crossing of energy eigenvalue curves)  may be accidental or may arise due to some symmetry of Hamiltonian. In one dimension it can be shown (see Appendix 1) that irrespective of whether the Hamiltonian is real or complex a degeneracy can not occur.

In this Letter, we study both the branches $[E^n_{\pm}(V_2)]$ together by fixing $V_1$ and varying $V_2$ (in the Abstract above 
we have denoted $V_2$ as $\lambda$) slowly and smoothly to reveal the accidental crossing of two energy eigenvalues curves unlike the Hermitian case in one dimension However, we will show the corresponding eigenstates become linearly dependent to dislodge degeneracy in one dimension, once again. We demonstrate some more features of eigenvalues including the coalescing of these two branches of real discrete eigenvalues to complex conjugate pairs when PT-symmetry breaksdown spontaneously (when $|V_2|>V_1+1/4$ [7]). 

In Schr{\"o}dinger equation, we generally take $2\mu=1=\hbar^2$.
Let us define [7]
\begin{equation}
s=\sqrt{1/4+V_1+|V_2|} \quad \mbox{and} \quad t=\sqrt{1/4+V_1-|V_2|},
\end{equation}
then the real discrete spectrum of Scarf II is given as 
\begin{equation}
E^n_{\pm}(t,s)=-(n-(s \pm t-1)/2)^2, \quad  n = 0,1,2,...\le [(s \pm t-1)/2],
\end{equation}
the square bracket $[\nu]$ denotes integer part of $\nu$..
Here, for a fixed $n$, $E^n_+ < E^n_-$ so we shall be calling $E_+$ as lower eigenvalue i.e. $E^0_+$ is the ground state. Unlike the case of real Hermitian Hamiltonians in the case of complex PT-symmetric 
Hamiltonians the oscillation theorem connecting the nodes of the eigenfunction with the quantum number $n$
does not follow. The respective eigenstates are given as [7] 
\begin{equation}
\psi^n_{\pm}(x,t,s)=A_{\pm} ~(\mbox{sech} x)^{(s \pm t-1)/2}~ \exp[-i\frac{1}{2}(s \mp t) \tan^{-1}(\sinh x)]~ P^{\mp t,-s}_n(i\sinh x).
\end{equation}
The Jacobi polynomials $P^{a,b}_n(i \sinh x)$ [15] are polynomials of degree $n$ in $\sinh x$ at large values of $|x|$ their divergence is damped  by $(\mbox{sech} x)^\nu$ as $n<\nu=(s \pm t-1)/2$. The exponential term is always finite. Consequently, we get $L^2-$integrable eigenfunctions satisfying Dirichlet
boundary condition.
  
Another interesting form of the eigenstates of (1)
can be written as [7]
\begin{equation}
\psi^n_{\pm}(x,t,s)= B_{\pm}~ (1-z)^{\mp t/2+1/4} (1+z)^{-s/2+1/4} P_n^{\mp t, -s}(z),\quad z=i\sinh x,
\end{equation}
which helps in the sequel.
These two branches ($\pm$) are kept in isolation. Let us see whether and when the condition  
\begin{equation}
E^m_{+}(t,s) = E^n_{-} (t,s), \quad m>n
\end{equation} 
is met. Using Eqs.(2,3) in (6)  the interesting condition is 
\begin{equation}
t=m-n, \quad m > n, 
\end{equation}
Thus, when $t(V_1,V_2)$ is an integer the two eigenvalue curves can intersect.
This in turn  means whenever $|V_2|$ equals the special values
\begin{equation}
V_{2*}=V_1+1/4-(m-n)^2, \quad  m>n, \quad m=1,2,3,..
\end{equation}
the eigenvalue curves $E^m_{+}(V_2)$ and $E^n_{-}(V_2)$ would cross each other. This seems to suggest an occurrence of degeneracy, however, we are in one dimension (see Appendix 1). Here we utilize an interesting rare property [16] of Jacobi polynomials, namely
\begin{equation}
P^{-j,s}_{n}(z)=C ~ (1-z)^{j}~ P^{j,s}_{n-j} (z), ~~j,n \in I^++\{0\},~~ j < n, ~s \in {\cal R},
\end{equation}
we prove this in the Appendix 2. Here, $I^+$ denotes the set of positive integers.
Therefore when $t=j$ we have 
\begin{equation}
E^{n-j}_-= E^n_+, \quad \psi^{n-j}_-(x,j,s) = C~ \psi^n_+(x,j,s),
\end{equation} 
where $C$ is independent of $x$ or $z$.
Hence, two intersecting energy eigenvalues will have corresponding eigenstates as linearly dependent, confirming no degeneracy in one dimension even in non-Hermitian PT-symmetric quantum mechanics.

When 
\begin{equation}
|V_2|>V_1+1/4 = V_{2c}, 
\end{equation}
PT-symmetry is spontaneously broken, $t$ becomes purely imaginary, i.e. $t=ir$
and all $E^n_{\pm}$ pairs become complex-conjugate.
\begin{equation}
E^n_{\pm}=-(n-(s\pm ir -1)/2)^2, r=\sqrt{|V_2|-V_1-1/4}~ \in {\cal R}, ~~ n=0, 1,2,... \le[(s-1)/2].
\end{equation}
The corresponding eigenstates are [7]
\begin{equation}
\psi^n_{\pm}(x)=D_{\pm} ~ (\mbox{sech} x)^{(s \pm ir-1)/2}~ \exp[\frac{1}{2}(-is \mp r) \tan^{-1}(\sinh x)]~ P^{\mp ir,-s}_n(i\sinh x).
\end{equation}
Consequently
\begin{equation} 
E^n_{+} = E^n_{-} \quad  \mbox{and} \quad \psi^n_{+}(x)  =\psi^n_{-}(x) \quad \mbox{when} \quad |V_2|=V_{2c}~ (r=0).
\end{equation} 
These eigenstates are $L^2$-integrable as the divergence of the Jacobi polynomial $P_n^{c,d}(i\sinh x)$ for large values of $|x|$ is suppressed by $(\mbox{sech} x)^{(s-1)/2}$ as 
$n < (s-1)/2$. Further, the argument of the exponential term is always finite irrespective of the values of $s,r$. Consequently the eigenstates vanish
asymptotically satisfying the Dirichlet boundary condition that $\psi(\pm \infty)=0$. 
Now we can clearly see that the eigenstates flip under the action of PT as [8,17,18,27,32]
\begin{equation}
(PT) \psi^n_{\pm}(x)= \psi^n_{\mp}(x).
\end{equation}
For $V_1=20$ and $V_2=0$, Scarf II is real Hermitian and it has 4 real discrete eigenvalues with states of definite parity: even/odd. Next we vary $V_2$ from 0 to $V_2 (>V_1)$. In Fig. 1(a) see the variation of $E^0_{\pm}(V_2)$, $E^0_+$ starts from  the eigenvalue of the real potential (1) i.e., $E^0=-16 $, when it is complex with $V_2\sim 5$ onwards $E^0_-$ starts existing then at $V_2=V_{2c}=20.25$ they both coalesce into one pair of complex conjugate eigenvalues (see the straight line ) demonstrating the spontaneous breakdown of PT symmetry. The rectangle from $V_2=0$ to $\sim 5$ is not a part of the curve it only denotes non-existence of any eigenvalue  ($E^0_-$). The linear part from $V_2=V_{2c}$  onwards is the variation the discrete real part of the complex conjugate pair of eigenvalues. Similarly, coalescing of $E^1_{\pm}$ and $E^2_{\pm}$ to complex conjugate pairs is shown in Figs.  1(b) and 1(c), respectively. Only $E^3_+$  exists and $E^3_-$ does not exist (see Fig. 1(c)).

In Fig. 2(a) all the curves of Fig. 1(a,b,c,d) are put together to present the hitherto un-noticed feature of
level crossings when $|V_2|$ takes values special values namely, $V_{2*}= 11.25, 16.25$ and 19.25 (see Eq. (8)). Fig. 2(b) shows the variation of $t(V_1=20,V_2)$ and at special values of $|V_2|=V_{2*}$ it becomes 3, 2 and 1. Since
discrete levels exist only for $n \le 3$, so we do not consider the case of $t=4$. Also see the special pairs of  eigenvalues becoming identical  in Table I at these special values of $|V_2|=V_{2*}$. See the claimed level crossings in Table I. For $|V_2|=11.25$ ( $t=2$) , we get just one level crossing
as we have $E^0_-=E^3_+$. For $|V_2|=16.25$ ($t=2$) giving two level crossings: $E^0_-=E^2_+$ and $E^1_-=E^3_+.$ When $|V_2|=19.25$ ($t=1$), notice three level crossings in Table I as we have $E^0_-=E^1_+;~ E^1_-=E^2_+;~ E^2_-=E^3_+.$ The eigenstates corresponding to these identical levels are linearly dependent as per Eq. (9).
\begin{figure}[ht]
\centering
\includegraphics[width=7 cm,height=5. cm]{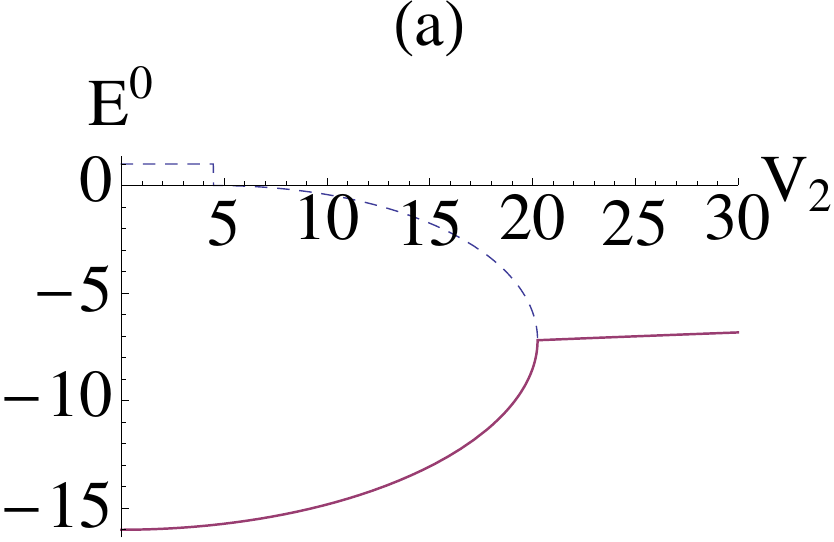}
\hskip .5cm
\includegraphics[width=7 cm,height=5. cm]{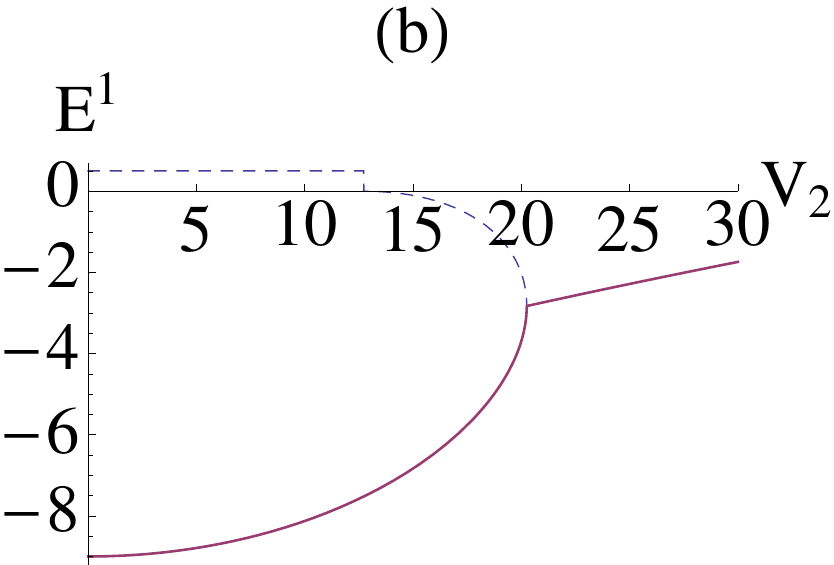}
\includegraphics[width=7 cm,height=5. cm]{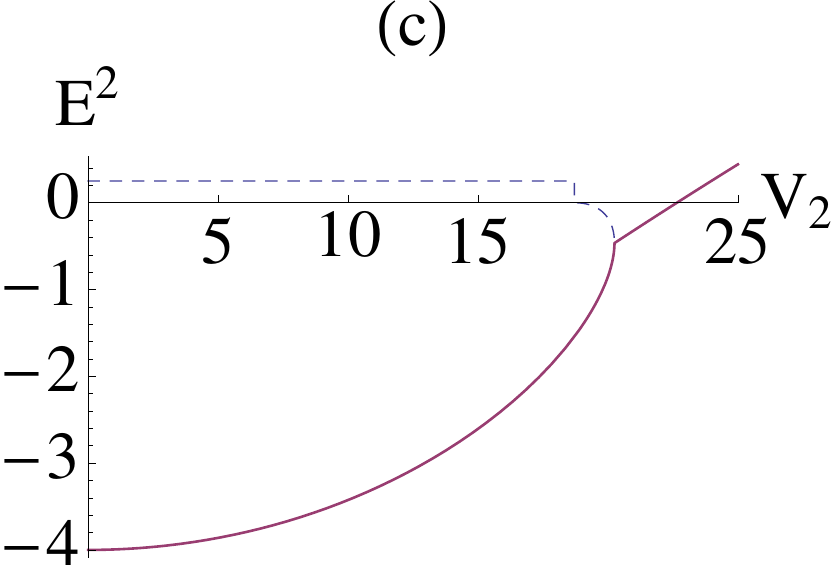}
\hskip .5cm
\includegraphics[width=7 cm,height=5. cm]{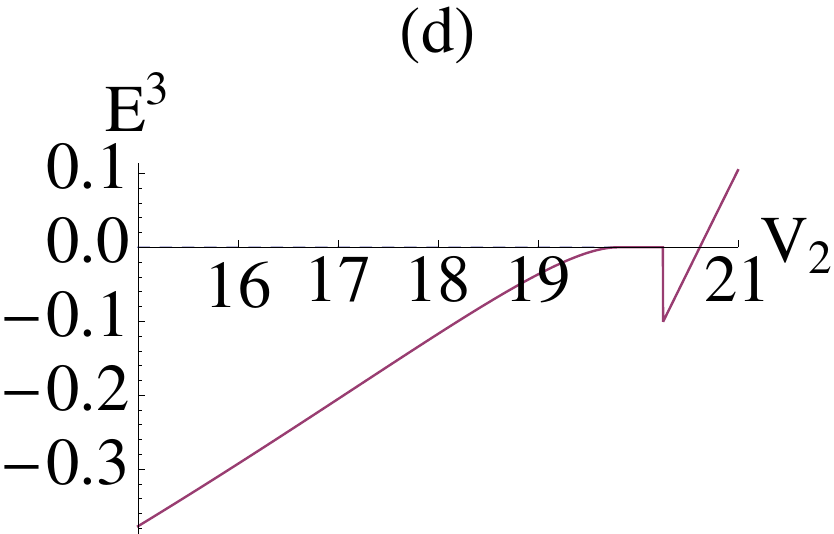}
\caption{Variation of real part of discrete eigenvalues, $E^n_\pm(V_2)$, of Scarf II (1) for $V_1=20$. The rectangles are not the part of the curves they just indicate absence of eigenvalues inside them. The curved parts display real eigenvalues and the linear parts the complex conjugate pairs starting from the critical value of $V_{2c}=20.25$. Solid (dashed) lines indicate +ve(-ve) branches.} 
\end{figure} 
\begin{figure}[ht]
\centering
\includegraphics[width=7 cm,height=5. cm]{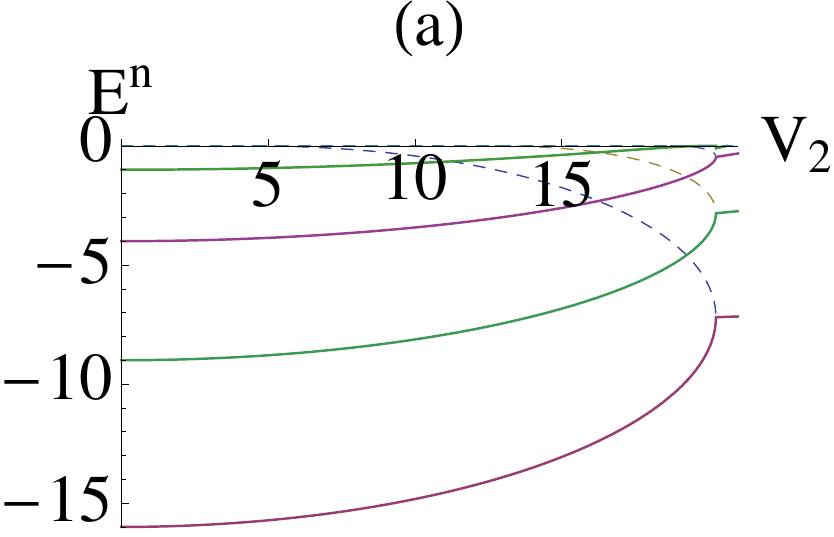}
\hskip .5cm
\includegraphics[width=7 cm,height=5. cm]{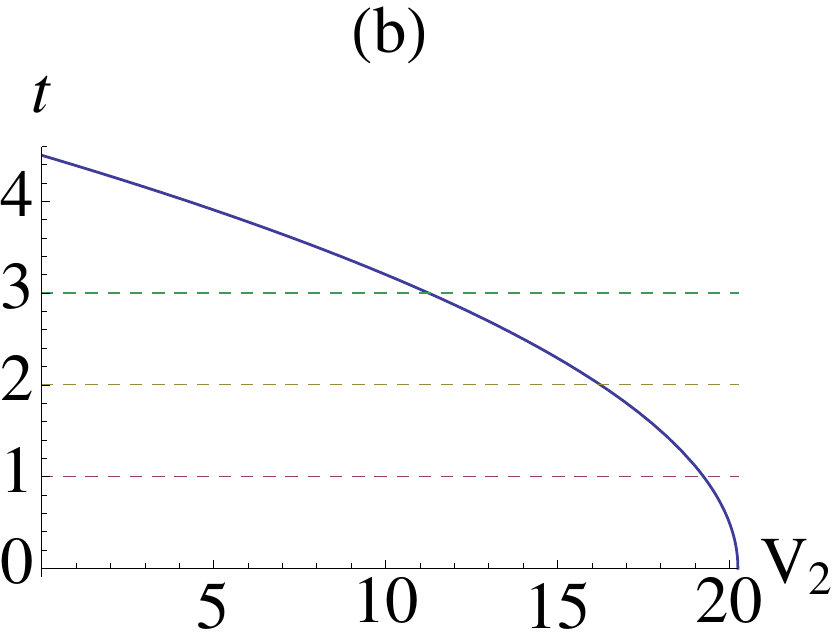}
\caption{(a): all the eigenvalue curves of Fig. 1 have been brought together, see several level crossings.  These crossings are accidental which occur at $|V_2|= V_{2*}= 11.25, 16.25, 19.25$ (see Eq. (8)) when (b): $t(20,V_2)$ takes integer values 1,2,3,..; $t=0$ indicating multiplicity of eigenvalues before real discrete eigenvalues coalesce to complex conjugate pairs at $|V_2|=V_{2c}=20.25$.} 
\end{figure}
\begin{figure}[ht]
\centering
\includegraphics[width=7 cm,height=5. cm]{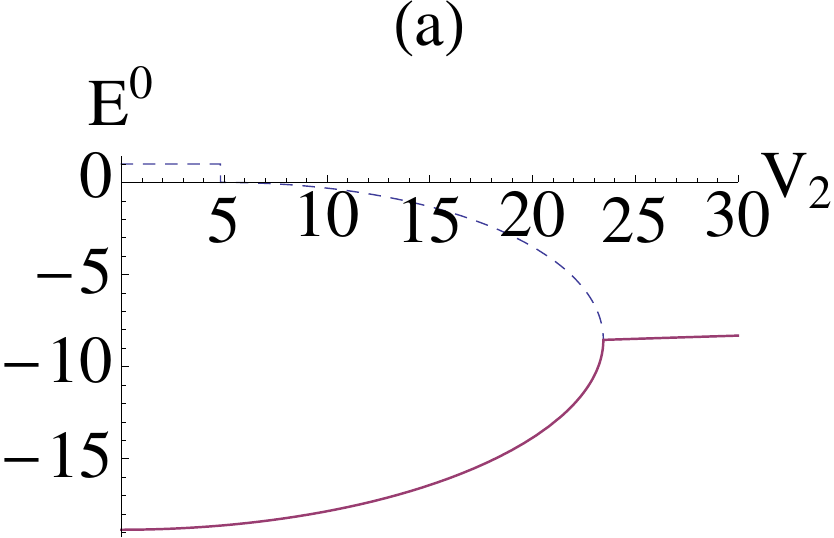}
\hskip .5cm
\includegraphics[width=7 cm,height=5. cm]{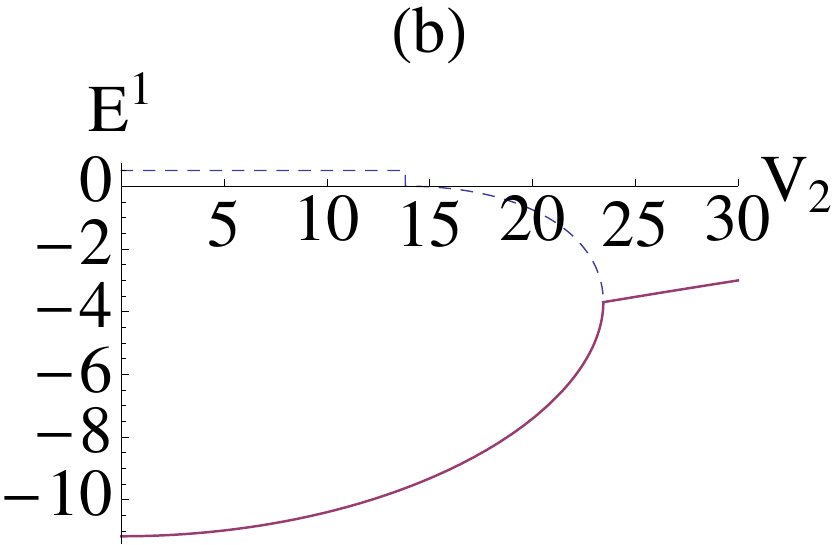}
\includegraphics[width=7 cm,height=5. cm]{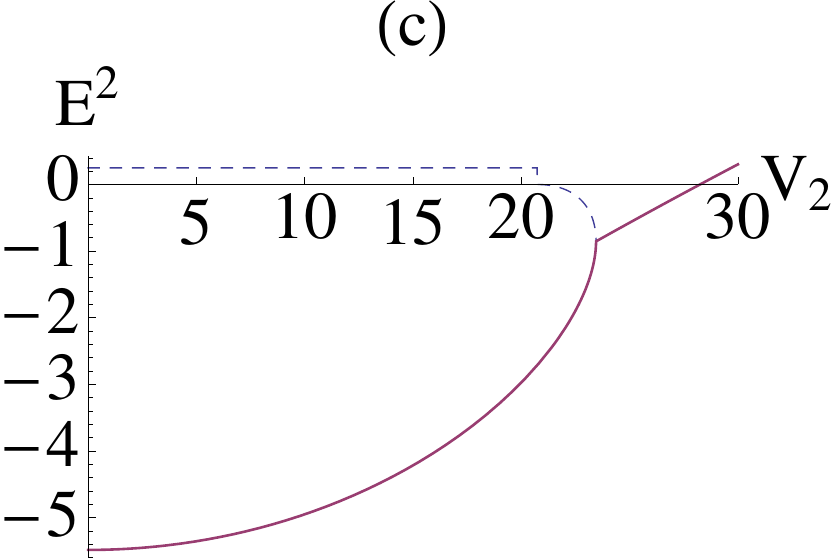}
\hskip .5cm
\includegraphics[width=7 cm,height=5. cm]{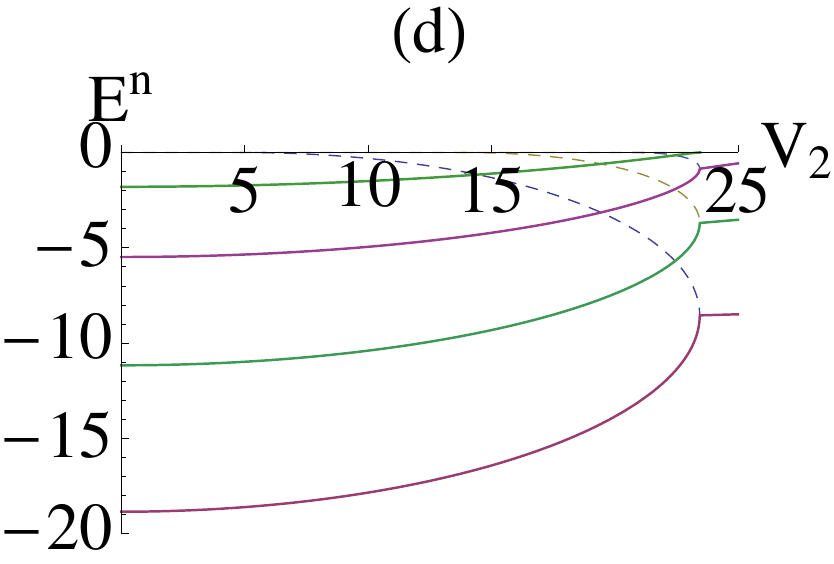}
\caption{The same as Fig. 2, for  a non-special value of $V_1=23.20$. Here the exceptional values are $V_{2c}=23.45$ and $V_{2*}= 14.45, 19.45, 22.45.$} 
\end{figure} 

\begin{figure}[ht]
\centering
\includegraphics[width=14 cm,height=6. cm]{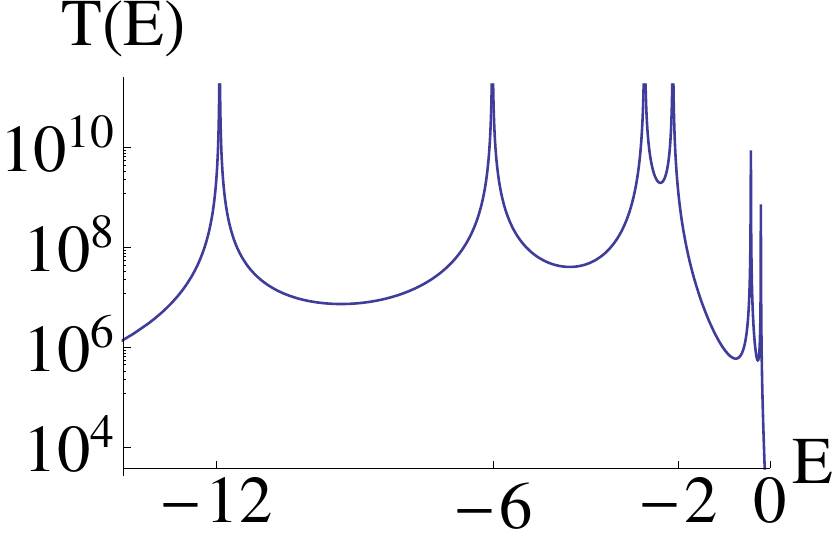}
\caption{The variation of transmission probability, $T(E)$, for negative energies for scarf II (1) when $V_1=20$ and $|V_2|=17$. All six negative energy poles correspond to six real discrete eigenvalues listed in Table I as: -11.92, -6.01, -2.72, -2.11, -0.42, -0.20.} 
\end{figure}
In Table I, the discrete eigenvalues of scarf II (1)
are given for a fixed value of $V_1=20$.
When $V_2=0$, the Hermitian potential (1) has four real discrete eigenvalues. Even when strength of the imaginary part i.e., $|V_2|=5$ we have four real eigenvalues only in the +ve branch. When $|V_2|>5$ the other branch starts picking up for instance when $|V_2|=9$ we have $E^0_-$ appearing; then on, the -ve branch starts developing.

\begin{table}[ht]
\caption{Discrete energy eigenvalues of Scarf II (1)
for $V_1=20 (2\mu=1=\hbar^2)$ $V_2$ being varied. Dash signs  mean absence of a discrete eigenvalue. At the crossings of levels $t(V_2)$ becomes non-negative integer irrespective of the values of $s(V_2)$
which are mostly non-integral.}
\begin{ruledtabular}
\begin{tabular}{|c||c||c||c||c||c||c||c||c||c||c|}
		\hline
$|V_2|$ & $t(V_2)$ & $s(V_2)$ & $E^0_-$ & $E^0_+$ & $E^1_-$ & $E^1_+$ & $E^2_-$ & $E^2_+$ & $E^3_-$ & $E^3_+$ \\
\hline	
0.0 & 4.5& 4.5  & $\--$ & -16.00 & $\--$ & -9.00 & $\--$ & -4.00 & $\--$ & -1.00 \\ \hline
5.0 & 3.90& 5.02 & $-0.003$ & -15.82 & $\--$ &- 8.86 & $\--$ & -3.91 & $\--$ & -0.95 \\ \hline
9.0 & 3.35& 5.40 & -0.27 & -15.06 &  $\--$ & -8.30 & $\--$ &-3.53 & $\--$ & -0.77 \\ \hline
11.25 & 3& 5.69 & -0.65 & -14.48 & -0.50& -7.87 & $\--$ &
-3.26 & $\--$ & -0.65	\\ \hline
16.25 &
2 & 6.04& -2.31 & -12.39 & -0.27 & -6.35 & $\--$ & -2.31 & $\--$ &
-0.27 \\ \hline
17 & 1.80 &6.10 & -2.72 & -11.92 & -0.42 & -6.01 & $\--$ & -2.11 & $\--$ & -0.20 \\ \hline
19.25 & 1 & 6.28 & -4.59 & -9.87 & -1.30 & -4.59 & -0.02 & -1.30 & $\--$ & -0.02 \\ \hline
20.25 & 0 & 6.363 & -7.19 & -7.19 & -2.82 & -2.82 & -0.46 & -0.46 & $\--$ & $\--$ \\ \hline 
20.30 & $0.22 i$ & 3.367  & -7.19 & -7.19 & -2.82 &-2.82  & -0.45  & -0.45 & $\--$ & $\--$ \\ 
& & & $-.60 i$ & $+.60 i$ & $-0.37 i$ & $+0.37 i$ & $-0.15 i$ & $-15 i$ & $\--$ & $\--$ \\ \hline 
21 & $0.86i$ & 6.42 & -7.16 & -7.16 & -2.74 & -2.74 & -0.31 & -0.31 & $\--$ & $\--$ \\
& & & $-2.34 i$ & $ +2.34 i$ & $-1.48 i$ & $1.48 i$ &
$-0.61 i$ & $0.61 i$ & $\--$ & $\--$ \\ \hline 
25 & $ 2.17i$ & 6.72 & -7.01 & -7.01 & -2.28 & -2.28 & $\underline{+0.44}$ & $\underline{ +0.44}$ & $\--$ & $\--$ \\
& & & $-6.24 i$ & $+6.24 i$ &$ -4.06i$ & $+4.06 i$ & $-1.88 i$ & $+1.88 i$ & $ \--$ & $\--$ \\ \hline  
\end{tabular}
\end{ruledtabular}
\end{table}

\begin{table}[ht]
\caption{The same as Table I, for non-special value of $V_1=23.20$.}
\begin{ruledtabular}
\begin{tabular}{|c||c||c||c||c||c||c||c||c||c||c|}
		\hline
$|V_2|$ & $t(V_2)$ & $s(V_2)$ & $E^0_-$ & $E^0_+$ & $E^1_-$ & $E^1_+$ & $E^2_-$ & $E^2_+$ & $E^3_-$ & $E^3_+$ \\
\hline	
0 &  4.84 & 4.84 & $\--$ & -18.85 & $\--$ & -11.17 & $\--$ & -5.48 & $\--$ & -1.80 \\ \hline
14.45 & 3 & 6.15 & -1.16 & -16.63 & -.006 & -9.47 & $\--$ & -4.31
& $\--$ & -1.16 \\ \hline
19.45 & 2& 6.54 & -3.15 & -14.25 & -.600 & -7.70 & $\--$ & -3.15 &
$\--$ & -.600 \\ \hline
22.45 & 1 & 6.77 & -5.70 & -11.47 & -1.92 & -5.70 & -0.15 & -1.92 & $\--$ & -0.15 \\ \hline
23.45 & 0 &  6.84 & -8.55 & -8.55 & -3.70 & -3.70 & -0.85 & -0.85 & $\--$ &  $\--$ \\ \hline
25 & $1.24 i$& 6.96 & -8.49 & -8.49 & -3.53 & -3.53 & -0.57 & -0.57 & $\underline{0.38}$ & $\underline{0.38}$ \\
& & & $-3.71 i$ & $3.71 i$ & $-2.46 i$ & $2.46 i$ & $-1.22 i$ &
$1.22 i$ & $-0.02i$ & $0.02 i$ \\ \hline
\end{tabular}
\end{ruledtabular}
\end{table}
Had the Scarf II not been solved analytically, then by numerical integration
of Schr{\"o}dinger equation for instance for  $V_2=17$ one would have got the energy eigenvalues as $-11.92, -6.01, -2.72, -2.11,-0.42, -0.20$ (see Table I) in the ascending order even without  knowing whether they belong to +ve or -ve branch. By virtue of exact analytic results (2-6) of Scarf II (1), we are familiar with two spectral branches of it. For example, we confirm
these eigenvalues as negative energy poles (Fig. 4) in the transmission probability, $T(E)$ [19] of Scarf II (1). The sequence of six eigenvalues is like : $E^0_+, E^1_+, E^0_-, E^2_+, E^1_-, E^3_+$ which appears arbitrary with regard to  subscripts and superscripts. We emphasize that there needs to be some theoretical basis under which we should know the number of spectral branches for a given complex PT-symmetric potential regardless of whether it is solved analytically or numerically. Next, all the branches  need to be mixed for full discrete spectrum. This would further require a new scheme to re-label the eigenvalues and  eigenstates.

When $V_2=20.25=V_1+.25$, all the energy doublets coalesce and become equal. Three equal pairs of
real discrete energies can be seen in Table I. Not shown here is the plot of $T(E)$ for this case which has three clear
poles at three negative energies: -7.49, -2.82, -0.46.
For $V_2=20.10$ there will be three closely lying doublets.

When $V_2>20.25$ all the eigenvalues become complex conjugate pairs  and their real part is shown to vary linearly in (a,b,c,d) parts. In Fig. 1(c) the real part of CCE (Complex Conjugate Eigenvalue)  also becomes positive but the corresponding eigenstates remain square integrable boundstates. See Table I,  for $V_2=21$, we find three pairs of CCE. For $V_2=25$, we find 3 pairs of CCE, the last pair is interesting as its real part is positive. It turns out that the eigenstates which are controlled by $n\le[((s-1)/2]$  will remain $L^2$-integrable bound states irrespective of whether real part of CCE is positive or negative. Apart from these real or complex-conjugate discrete states the Scarf II will have continuous part of spectrum where, in various parametric regimes, it  displays interesting novel scattering properties [20,21] like non-reciprocity [22], spectral singularity [23], coherent perfect absorption with [24] and without [25] lasing.

We would like to re-emphasize that it is $t(V_2)$ (2) which becomes a non-negative integer irrespective of the value of $s(V_2)$ which are usually non-integral or non-real (See the Table I). Since, in Eq. (1) $V_1=20= 4 \times 5$ is of a special type $m(m+1), m=1,2,3,..$ for which the real Scarf II potential becomes reflectionless. Hence, the level-crossings may be (mis)taken as arising only due to the reflectionlessness of the real part of the potential. Therefore, we consider a non-special value of $V_1=23.20$ and present results in Fig. 3 and Table II.

When PT-symmetry breaks  down spontaneously as a parameter of the potential passes over a critical value, each of these pairs change over to CCE. Before and after this transition the eigenstates would be orthogonal in the proposed way [26]
\begin{equation}
\int_{-\infty}^{\infty} \psi_1(x) \psi_2(x)~dx=0, \quad E_1 \ne E_2.
\end{equation}
We conclude that in the evolution of eigenvalues as the imaginary strength ($V_2$) of the potential is varied slowly 
there occur two types of special values (exceptional : EP) $V_{2c}$ (11) and $V_{2*}$ (8) of $V_2$ at  which two branches (3) of eigenvalues (coincide) coalesce (10,14) but the corresponding eigenstates are identical or linearly dependent. In the former case,  real eigenvalues go over to complex conjugate eigenvalues and in the latter case, real eigenvalues cross over to real eigenvalues. The former case is  well known as spontaneous breaking of PT-symmetry,  the latter  is much rarer and a unique instance in the PT-symmetric quantum mechanics, so far; we call it accidental crossing of eigenvalues. As shown here in Figs. (1) and (2) and in the Tables I and II, for Scarf II when $V_1=20$, $V_{2c}=20.25$ (only one value), whereas $V_{2*}= 11.25, 16.25, 19.25$  (8). When  $V_1=23.20$, we have  one value of $V_{2c}=23.45$ but three values of  $V_{2*}=14.45, 19.45, 22.45$. In general there may be more than one (exceptional) points of the type $V_{2c}$ for a complex PT-symmetric potential [1,18]). 

Surprisingly, earlier [27] the parametric point $\lambda_c$ where two eigenvalues coalesce (coincide) have been termed as a point of non-Hermitian degeneracy without realizing that corresponding eigenfunctions become linearly dependent as revealed here in Eq. (10) for Scarf II. Similarly, the accidental crossing of real eigenvalues could be sensed but its presence (absence)  has been discussed [28] in terms degeneracy  (non-degeneracy) in one-dimension! 

The exceptional points $V_{2c}$ and $V_{2*}$ of a PT-symmetric complex potential are the parametric points  where the Hamiltonian becomes non-diagonalizable. For instance, the following two matrices having coincident eigenvalues are non-diagonalizable
\begin{equation}
A= \left (\begin{array}{cc} 1 & 0 \\1 & 1\end{array} \right),
\psi_1=\psi_2 = \left (\begin{array}{c} 0  \\ 1 \end{array}
 \right); \quad A= \left (\begin{array}{ccc} 1 & 1 & 0 \\0 & 1 & 1 \\ 0 & 0 & 1 \end{array} \right), \quad \psi_1= \psi_2=\psi_3= \left (\begin{array}{c} 1  \\ 0  \\ 0 \end{array}
 \right).
\end{equation}
As per the characteristic equation $\det|A-\lambda I|=0$ these matrices have 1 as repeated eigenvalue with only one eigenvector
which alone can not constitute an invertible diagonalizing matrix $D$ such that $D^{-1} A D= \Lambda.$  Notably these matrices are pseudo-Hermitian under some metric $\eta$  as $\eta^{-1} A \eta =A^\dagger$. In contrast to the pseudo-Hermitian matrix, a Hermitian matrix is always diagonalizable. The PT-symmetric Hamiltonian $H=-\frac{d^2}{dx^2}+V(x)$ are P-pseudo-Hermitian and  a loss of diagonalizability  of $H$ at a critical (exceptional) point wherein  PT-symmetry is broken has already been discussed [29] as a possibility (not a necessity). This can be demonstrated in a simple way as below [30].
\begin{equation}
H= \left (\begin{array}{cc} a+b & ic \\ic & a-b\end{array} \right),
\eta= \left (\begin{array}{cc} 1 & 0 \\0 & -1 \end{array} \right),
 E_{1,2}=a\pm\sqrt{b^2-c^2}, \psi_{1,2}= \left (\begin{array}{c} i(b\mp\sqrt{b^2-c^2})/c \\ 1 \end{array}\right)
\end{equation}
Let $a,b,c$ be real, then $H$ is pseudo-Hermitian under the metric $\eta$. When $c$  critically equals $b$ both eigenvalues coincide and eigenvectors become identical or linearly dependent rendering $H$ as non-diagonalizable.

As per the discussion in the Appendix 1 or Kato's [31] theory of exceptional points, the crossing of eigenvalues in one-dimensional complex PT-symmetric potentials may not be unexpected. Nevertheless, the fact remains that among several  models [1,2,4,13,18] only complex Scarf II (1)  exhibits this phenomenon (See Figs. 2(a), 3(d)), so far. In non-Hermitian  (PT-symmetric and Pseudo-Hermitian) quantum mechanics several novel phenomena [20-25] occur as a possibility but not as as a necessity. For instance, the potential like $x^2+igx$ exhibit neither spontaneous breakdown of PT-symmetry nor the acclaimed crossing of real discrete eigenvalues.

It may be recalled here that  ``accidental degeneracies" in hydrogen atom have been explained by Pauli in terms of the invariance of the  Laplace-Runge-Lenz vector. In the same vein, the acclaimed ``accidental level crossings" may not be so accidental, if the presence  of Lax-Novikov integral [32]  in such a potential system as Scarf II (1) is studied.  Next, one may wonder as to what kind of complex PT-symmetric potential may exhibit the level crossings. Some very interesting properties of complex Scarf II discussed in Refs. [33] may be helpful. Investigations in this regard are welcome.

We would like to summarize that  if one isolates  two branches of discrete eigenvalues of complex PT-symmetric Scarf II, the coalescing of real eigenvalues to complex conjugate eigenvalue (CCE) pairs and hence the spontaneous breakdown of PT-symmetry can not be discussed. And if one brings these two branches together, interesting variations of discrete eigenvalues  emerge. In this, the most interesting is the accidental crossings of two eigenvalues. Since level crossing means degeneracy which can not occur in one dimension, we could succeed in showing the linear dependence of such pairs of eigenstates by proposing and proving a rare property of Jacobi polynomials.

Given a complex PT-symmetric potential regardless of whether it is solved numerically or analytically there is a need of a criterion to tell the number of spectral branches: one or more. Further, the spectral branches need to be mixed for full discrete spectrum. Furthermore, the next example of a complex one-dimensional PT-symmetric potential displaying accidental crossing of eigenvalues would be most welcome.  

Finally, we would like  re-affirm  that we have demonstrated novel exceptional (branch) points in the spectrum of complex PT-symmetric Scarf II potential, where two real eigenvalues coincide and cross over to real eigenvalues but the corresponding eigenstates are identical or linearly dependent. Though, this whole demonstration is entirely analytic yet figures and tables presented here make it even more transparent.
\section*{\Large{Appendix 1}}
\renewcommand{\theequation}{A-\arabic{equation}}
\setcounter{equation}{0}
\noindent
{\bf Proposition 1:}\\ \\
Let $\psi_m(x), \psi_n(x)$ be two $L^2-$integrable solutions of one dimensional time-independent Schr{\"o}dinger equation satisfying Dirichlet boundary
condition $\psi(\pm \infty)$ with and having equal eigenvalue $E$, then $\psi_m(x)$ and $\psi_n(x)$ are linearly dependent. 

{\bf Proof:}
Let the potential $V(x)$ (real or complex) in Schr{\"o}dinger $(2\mu =1=\hbar^2)$ equation gives rise to two solutions
$\psi_m(x)$ and $\psi_n(x)$ with the same energy eigenvalue $E$, then we write
\label{allequations}
\begin{eqnarray}
\frac{d^2\psi_m(x)}{dx^2}+[E-V(x)]\psi_m(x)=0, \\ 
\frac{d^2\psi_n(x)}{dx^2}+[E-V(x)]\psi_n(x)=0.
\end{eqnarray}
Multiply the first by $\psi_n(x)$ and the second by $\psi_m(x)$  and by subtracting them we get
\begin{equation}
\psi_m(x) \frac{d^2\psi_n(x)}{dx^2}-\psi_n(x) \frac{d^2\psi_m(x)}{dx^2}= 0 \Rightarrow \frac{d}{dx}\left (\psi_m \frac{d\psi_n}{dx}-
\psi_n \frac{d\psi_m}{dx}\right)=0,
\end{equation}
leading to
\begin{equation}
\left ( \psi_m(x) \frac{d \psi_n(x)}{dx}-
\psi_n(x) \frac{d \psi_m(x)}{dx} \right) = C,
\end{equation}
where $C$ is constant independent of $x$ which can as well be determined at $x=\pm \infty$. As the eigenstates satisfy $\psi_j(\pm \infty)=0$, we get $C=0$. Further we get, $\frac{\psi_m'(x)}{\psi_m} = \frac{\psi_n'(x)}{\psi_n}$ implying linear dependence: $\psi_m(x)= C' \psi_n(x)$. Thus, like in Hermitian quantum mechanics, here too the degeneracy can not occur.
\section*{\Large{Appendix 2}}
\renewcommand{\theequation}{A-\arabic{equation}}
\setcounter{equation}{0}
\noindent
One of the representations of Jacobi polynomials [15] is
\begin{equation}
P^{a,b}_n(z) = \frac{\Gamma(a+n+1)}{n! ~\Gamma(a+b+n+1)} \sum_{m=0}^n  {n \choose m} \frac{\Gamma(a+b+n+m+1)}{\Gamma(a+m+1)} \left(\frac{z-1}{2} \right)^m.
\end{equation}
Putting $a=-j,~b=s$ in the above expression we get
$$P^{-j,s}_n(z) = \frac{\Gamma(n-j+1)}{n! ~\Gamma(s+n-j+1)} \sum_{m=j}^n  {n \choose m} \frac{\Gamma(s+n+m-j+1)}{\Gamma(m-j+1)} \left(\frac{z-1}{2} \right)^m,$$
the summation, instead of $m=0$, will effectively start from $m=j$ as Gamma of zero and negative integers is $\pm \infty$. Changing the summation index to $p=m-j$ gives, 
$$P^{-j,s}_n(z) = \frac{\Gamma(n-j+1)}{n! ~\Gamma(s+n-j+1)} \sum_{p=0}^{n-j}  {n \choose {p+j}} \frac{\Gamma(s+n+p+1)}{\Gamma(p+1)} \left(\frac{z-1}{2} \right)^{p+j}.$$
Opening ${n \choose {p+j}}$ and manipulating the Gamma functions we get
\begin{equation}
P^{-j,s}_n(z) = \frac{(n-j)!}{(n-j)! ~\Gamma(s+n-j+1)} \sum_{p=0}^{n-j}  {{n-j} \choose p} \frac{\Gamma(s+n+p+1)}{\Gamma(p+j+1)} \left(\frac{z-1}{2} \right)^{p+j}.
\end{equation}
Using Eq.(A-1), to substitute for $P^{j,s}_{n-j}(z)$, in Eq.(A-2) we prove
\begin{equation}
P^{-j,s}_n(z) = (-2)^{-j}~ \frac{(n-j)!}{n!} ~\frac{\Gamma(s+n+1)}{\Gamma(s+n-j+1)} ~~(1-z)^j~~ P^{j,s}_{n-j}(z).
\end{equation}
\section*{\Large{Acknowledgement}} It gives me (ZA) a pleasure to acknowledge interesting discussions with Dhruv Sharma (NIT, Rourkela, India).

\end{document}